\documentclass[epj]{webofc}
\usepackage[varg]{txfonts}   

\usepackage{enumerate}
\usepackage{amsmath,amssymb}
\usepackage{mathrsfs}
\usepackage{bm}
\wocname{EPJ Web of Conferences}
\woctitle{CONF12}
\begin{document}
\selectlanguage{english}
\title{Gauge-independent Higgs mechanism and the implications for quark confinement
}
%
%

\author{Kei-Ichi Kondo\inst{1}\fnsep\thanks{\email{kondok@faculty.chiba-u.jp}}
}

\institute{Department of Physics,  Faculty of Science, Chiba University, Chiba 263-8522, Japan
}

\abstract{%
We propose a gauge-invariant description for the Higgs mechanism by which a gauge boson acquires the mass.
We do not need to assume spontaneous breakdown of gauge symmetry signaled by a non-vanishing vacuum expectation value of the scalar field. 
In fact, we give a manifestly gauge-invariant description of the Higgs mechanism in the operator level, which does not rely on spontaneous symmetry breaking. 
For concreteness, we discuss the gauge-Higgs models with $U(1)$ and $SU(2)$ gauge groups explicitly. 
This enables us to discuss the  confinement-Higgs complementarity from a new perspective.
}
\maketitle
\section{Introduction}
\label{intro}

\textbf{Spontaneous symmetry breaking (SSB)} is an important concept in physics.
It is known that SSB occurs when the lowest energy state or the vacuum is degenerate \cite{Nambu}.

First, we consider the SSB of the \textbf{global  continuous symmetry}  $G=U(1)$.
The complex scalar field theory described by the following Lagrangian density $\mathscr{L}_{\rm cs}$ 
  has the global $U(1)$ symmetry:  
\begin{align}
\mathscr{L}_{\rm cs} =   \partial_{\mu}\phi^{*} \partial^{\mu} \phi - V(\phi^{*} \phi) , \quad
V(\phi^{*} \phi) =  \frac{\lambda}{2} \left( \phi^{*} \phi - \frac{\mu^{2}}{\lambda} \right)^{2} , \ \phi \in \mathbb{C} , \ \lambda > 0 ,
\end{align}
where $*$ denotes the complex conjugate and the classical stability of the theory requires $\lambda > 0$.
Indeed, this theory has the global $U(1)$ symmetry, since $\mathscr{L}_{\rm cs}$ is invariant under the global $U(1)$ transformation, i.e., the phase transformation: 
\begin{align}
 \phi(x) \to e^{i\theta} \phi(x) .
\end{align}

\begin{figure}[h]
\centering
\includegraphics[width=4cm]{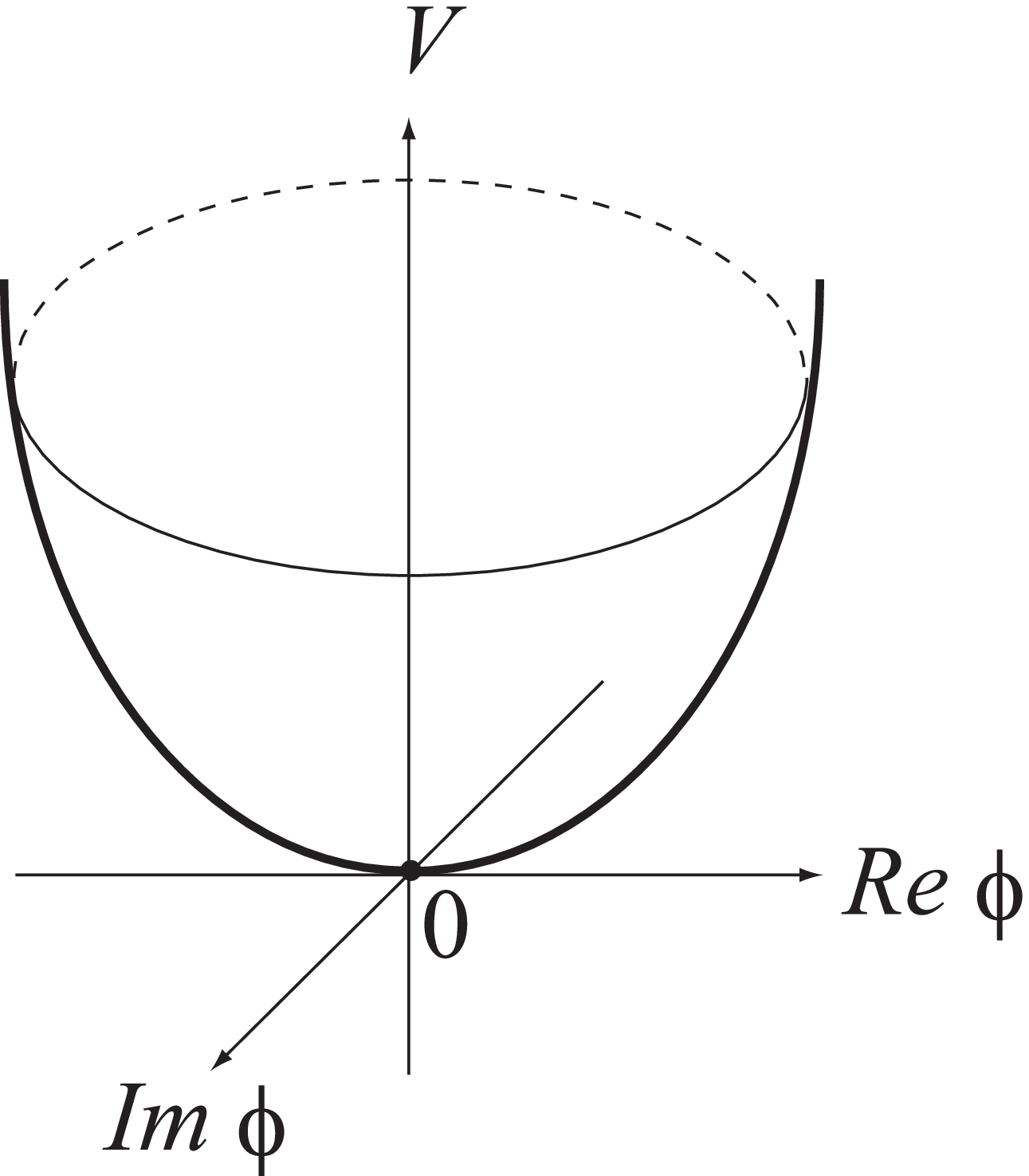}
\quad\quad
\includegraphics[width=4cm]{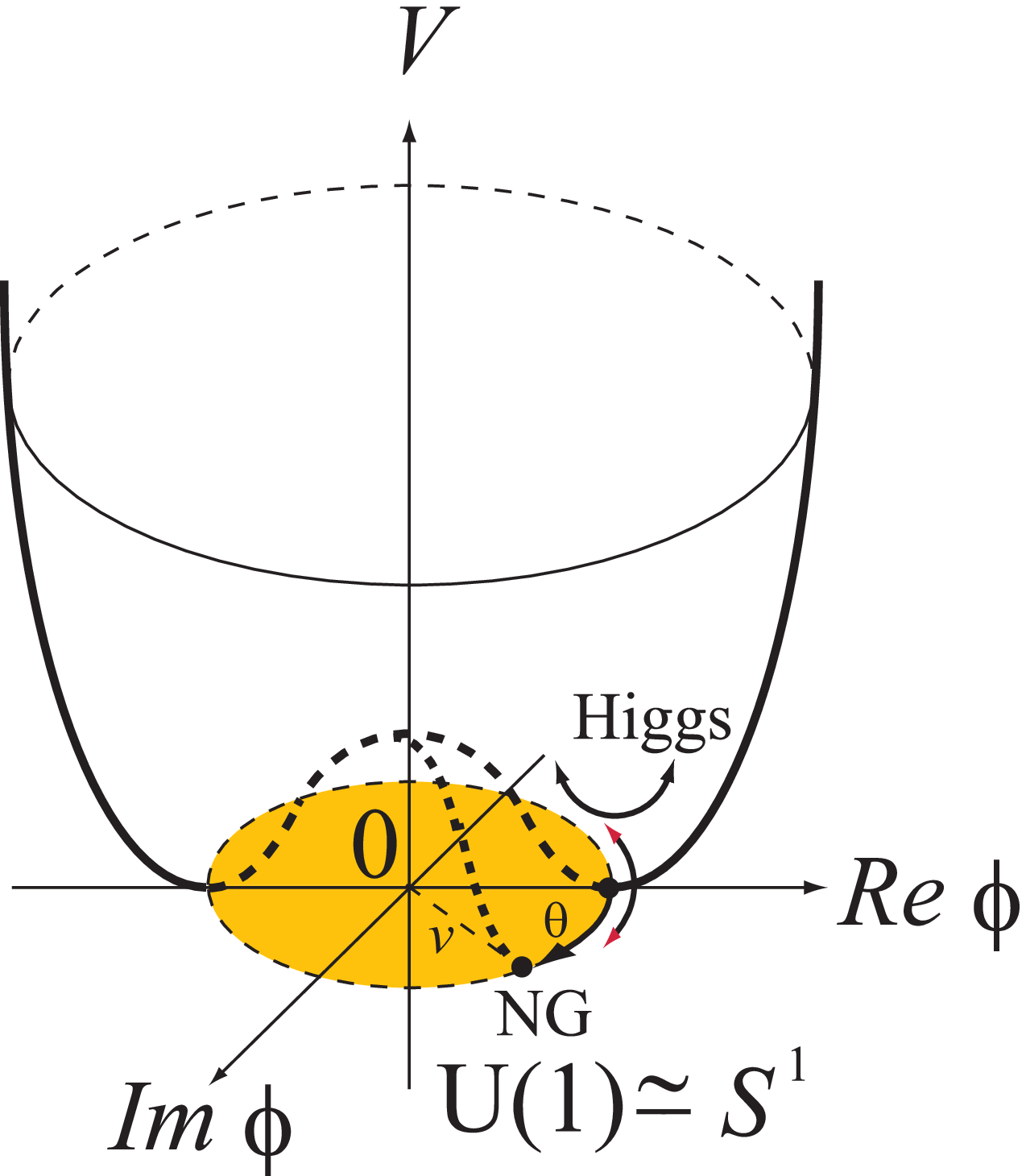}
\caption{
(Left) unbroken $U(1)$ symmetry   ($\mu^2 \le 0$):
$\langle \phi(x) \rangle =0$, 
(Right)
SSB of $U(1)$ symmetry ($\mu^2>0$): 
$\langle \phi(x) \rangle = \sqrt{\mu^2/\lambda}  \ne 0$
}
\label{fig:SSB-U(1)}   
\end{figure}


If the potential has the unique minimum ($\mu^2 \le 0$) as shown in the left panel of Fig.~\ref{fig:SSB-U(1)}, then the vacuum is unique and the $U(1)$ symmetry is unbroken: the scalar field has the vanishing \textbf{vacuum expectation value} (VEV):
$\langle \phi(x) \rangle =0$.
If the potential is modified to take the shape of the bottom of a wine bottle ($\mu^2 > 0$) as shown in the right panel of Fig.~\ref{fig:SSB-U(1)}, then the vacuum is degenerate and every point on the circle with the radius   $\sqrt{\mu^2/\lambda}$ can be a candidate of the vacuum.  Once one of the points on the circle is chosen as the vacuum, the $U(1)$ symmetry is spontaneously broken and the scalar field has the non-vanishing VEV:
$\langle \phi(x) \rangle = \sqrt{\mu^2/\lambda}  \ne 0$.

According to the Nambu-Goldstone theorem \cite{Nambu,Goldstone}, 
there appears the massless \textbf{Nambu-Goldstone particle} associated with the SSB, since $G=U(1)$ is a  continuous group. 
In fact, using the parameterization of the scalar field called the polar form:
\begin{align}
\phi(x) = |\phi(x)| e^{i\pi(x)/v} \in \mathbb{C} ,
\end{align}
we find the following identification:
\\
$\bullet$ $\pi(x)$ describes the flat direction along the circle: $\to $  massless   {Nambu-Goldstone mode}  
\\
$\bullet$ $|\phi(x)|$ describes the curved direction orthogonal to the flat direction: $\to $  massive  Higgs mode.

Next, we consider what happens in the gauge field theory with the \textbf{local continuous symmetry} $G=U(1)$.  
For this purpose, we consider the gauged complex scalar field theory, 
namely, \textbf{$U(1)$ gauge-scalar theory} with the Lagrangian density:
\begin{align}
\mathscr{L}_{\rm AH} =& - \frac{1}{4} F_{\mu\nu} F^{\mu\nu} +  ( D_{\mu}\phi)^{*} (D^{\mu}\phi) - V(\phi^{*} \phi) , \
 V(\phi^{*} \phi) =  \frac{\lambda}{2} \left( \phi^{*} \phi - \frac{\mu^{2}}{\lambda}  \right)^{2} , \ \phi \in \mathbb{C} , \ \lambda > 0 ,
 \nonumber\\
 & F^{\mu\nu} = \partial^{\mu} A^{\nu} - \partial^{\nu} A^{\mu} , \quad D_{\mu} = \partial_{\mu} - i q A_{\mu} ,
\label{L1}
\end{align}
where 
$F^{\mu\nu}(x)$ is the field strength for the $U(1)$ gauge field $A^{\mu}(x)$ 
and
 and $D_{\mu}$ is the covariant derivative 
 for the complex scalar field $\phi(x) \in \mathbb{C}$ with 
$q$ being the electric charge of $\phi(x)$. 
Indeed, this theory has the local $U(1)$ gauge symmetry, since $\mathscr{L}_{\rm AH}$ is invariant under the local gauge transformation: 
\begin{align}
 \phi(x) \to e^{iq\theta(x)} \phi(x) . 
 \quad
 A_\mu(x) \to e^{iq\theta(x)} [A_\mu(x) + iq^{-1} \partial_\mu ] e^{-iq\theta(x)} .
\end{align}

The $U(1)$ gauge symmetry is completely  broken spontaneously by choosing a vacuum with the non-vanishing VEV of the scalar field, e.g., $\langle \phi(x) \rangle = v/\sqrt{2}  \ne 0$. Then, the gauge-invariant kinetic term of the scalar field is modified to the form which includes the mass term for the gauge field $A_\mu$ (and the kinetic term of the scalar field plus interactions between the scalar field and the gauge field):
\begin{align}
 |D_{\mu}\phi|^2 = |\partial_{\mu}\phi - i q A_{\mu}\phi |^2  
\to |- i q A_{\mu}\langle \phi(x) \rangle |^2 + ... = \frac12 (qv)^2 A_{\mu}A^{\mu} + ... .
\end{align}
Consequently, the massless \textbf{Nambu-Goldstone particle} $\pi$ (boson) associated with the SSB of $U(1)$ symmetry disappears from the spectrum, since it is absorbed into the massless gauge boson $A_\mu$ to form the massive vector boson as  the longitudinal component. 
 This is a model for superconductor. 
The superconductivity is understood as a result of SSB of $U(1)$ gauge symmetry. The Meissner effect is just the Anderson-Schwinger mechanism. 
This is a special case of the \textbf{Brout-Englert-Higgs (Guralnik-Hagen-Kibble) mechanism} or \textbf{Higgs phenomenon} which is the most well-known mechanism by which gauge bosons acquire their masses \cite{Higgs1,Higgs2,Higgs3}.

However, the \textbf{spontaneous gauge symmetry breaking} is a rather misleading terminology.
Remember that the lattice gauge theory $\grave{a}\  la$ Wilson \cite{Wilson74} gives  a well-defined gauge theory \cite{YM54} without gauge fixing. 
The \textbf{Elitzur theorem} \cite{Elitzur75} tells us that  \textit{the  local continuous gauge symmetry cannot break spontaneously, if no gauge fixing  is introduced}. 
 In the absence of gauge fixing, all gauge non-invariant Green functions vanish identically.
Especially, the VEV $\langle \bm{\phi} \rangle$ of   $\bm{\phi}$ is rigorously zero regardless of the shape of the scalar potential $V$: 
\begin{align}
 \langle \bm{\phi} \rangle = 0 ,
\end{align}
 Therefore, we are forced to fix the gauge to cause the non-zero VEV. 
Even after the gauge fixing, however, we still have the problem. 
\noindent
\textbf{Whether SSB occurs or not depends on the gauge choice.}
In non-compact $U(1)$ gauge-Higgs model, the SSB occurs $\langle \bm{\phi} \rangle \ne 0$ only in the Landau gauge $\alpha=0$, and no SSB occur $\langle \bm{\phi} \rangle = 0$ in all other covariant gauges with $\alpha \ne 0$, as rigorously shown \cite{KK85,BN86}. 
In axial gauge, $\langle \bm{\phi} \rangle = 0$ for compact models \cite{FMS80}.


 After imposing the gauge fixing condition for the original local gauge group $G$, a global subgroup $H^\prime$ remains unbroken. 
Such a global symmetry  $H^\prime$ is called the \textbf{remnant global gauge symmetry} \cite{GOZ04,CG08}. 
\textit{Only a remnant global gauge symmetry $H^\prime$  of the local gauge symmetry $G$ can break spontaneously} to cause the Higgs phenomenon \cite{Kugo81}.
However, such subgroup $H^\prime$ is not unique  and the location of the breaking in the phase diagram depends on the remnant global gauge symmetries in the gauge-Higgs model. 
The relevant numerical evidences  are given on a lattice \cite{CG08} for different remnant symmetries allowed for various confinement scenarios. 
Moreover, the transition occurs in the regions where the Fradkin-Shenker-Osterwalder-Seiler theorem \cite{FS79,OS78}  assures us that there is no transition  in the phase diagram. 


 The above observations indicate that \textit{the Higgs mechanism should be characterized in a gauge-invariant way without breaking the original gauge symmetry.}
\noindent
[It is obvious that the non-vanishing VEV of the scalar field is not a gauge-invariant criterion of SSB.]
We show that \textit{a gauge boson can acquire the mass in a gauge-invariant way without assuming spontaneous breakdown of gauge symmetry which is signaled by the non-vanishing VEV of the   scalar field.}  
The Higgs phenomenon can be described even without such SSB. 
The spontaneous symmetry breaking is sufficient but not necessary for the Higgs mechanism to work. 

 Remember that {quark confinement is realized in the unbroken gauge symmetry phase with \textbf{mass gap}. }
Thus, the gauge-invariant description of the Higgs mechanics can shed new light on the \textbf{complementarity  between confinement phase and Higgs phase.}


\section{Gauge-invariant Higgs mechanism in U(1) case:   complete SSB}

We consider the {Abelian-Higgs theory} or  {$U(1)$ gauge-scalar theory} with the Lagrangian density (\ref{L1}).
For $\mu^2>0$, the minimum of the potential is attained when the magnitude of the scalar field is equal to the value: 
\begin{equation}
		|\phi(x)| = \frac{v}{\sqrt{2}}  , \quad v = \sqrt{\frac{\mu^2}{\lambda/2}} .
\end{equation}
We restrict our consideration to the radially fixed scalar field satisfying 
\begin{equation}
		 \phi^*(x) \phi(x)  =  \frac{v^2}{2}   ,
\end{equation}
to freeze the degree of freedom for the Higgs particle (The recovery of the Higgs particle will be discussed in the final section). 
If we use   the {polar representation} for the  radially fixed scalar field: 
\begin{align}
	&	\phi(x) = \frac{v}{\sqrt{2}} e^{i \pi(x)/v} \in \mathbb{C} , \quad \pi(x) \in \mathbb{R} ,
	\label{polar}
\end{align}
the covariant derivative reads 
\begin{align}
D_\mu \phi=(\partial_{\mu} - iq A_{\mu})\phi 
= - \frac{v}{\sqrt{2}} iq  \left( A_{\mu} - \frac{1}{qv} \partial_{\mu} \pi    \right) e^{i \pi/v} ,
\end{align}
and the kinetic term of the scalar field reads
\begin{align}
		(D_{\mu} \phi)^{*} (D^{\mu} \phi) 
=&   \frac12 (qv)^2 \left( A_{\mu} - \frac{1}{qv} \partial_{\mu} \pi   \right)^2 .
\end{align}
Therefore, one can introduce a new vector field $W_{\mu}$ defined by
\begin{equation}
		W_{\mu}(x) := A_{\mu}(x) - m^{-1} \partial_{\mu} \pi(x) , \ m := qv, 
		\label{Proca}
\end{equation}
so that $\mathscr{L}_{\rm AH}$ is completely rewritten in terms of   $W_{\mu}$:
\begin{align}
 \mathscr{L}_{\rm{AH}} = -\frac{1}{4} (\partial_{\mu} W_{\nu} - \partial_{\nu} W_{\mu})^2 + \frac{1}{2} m^2 W_{\mu} W^{\mu} . 
\label{L2}
\end{align}
The field $\pi$ is usually interpreted as the massless Nambu-Goldstone boson  associated with the complete SSB $G=U(1) \to H=\{ 1 \}$, which is absorbed into the massive field $W_{\mu}$ to be identified with the longitudinal component of the massive gauge boson.  

However, we find that the massive vector field $W_{\mu}$ has a manifestly gauge-invariant representation written in terms of the original scalar field $\phi$ and the gauge field $A_\mu$ \cite{Kondo16}: 
\begin{equation}
 W_{\mu}(x)   
= iq^{-1} \hat{\phi}^*(x)  D_\mu \hat{\phi}(x) 
= -iq^{-1}  \hat{\phi}(x)  D_\mu \hat{\phi}^*(x)  ,
\quad
( \hat{\phi}(x):=\phi(x)/|\phi(x)| )  .
\label{U1}
\end{equation}
Substituting (\ref{U1}) into (\ref{L2}) reproduces the original Lagrangian  (\ref{L1}).
In particular, it is easy to check that (\ref{U1}) reduces to (\ref{Proca}) for the parameterization (\ref{polar}). 
\textit{The representation (\ref{U1}) for the massive vector field  gives a parameterization-independent description of the Higgs phenomenon and does not rely on the SSB of the gauge symmetry}.
Therefore, the Higgs mechanism can be described using  the other  parameterizations, e.g., the decomposition of the complex field into the real and imaginary parts,
\begin{align}
	\phi(x) = \frac{1}{\sqrt{2}} [v+\varphi(x) + i\chi(x)] .
\end{align}

It is shown that 
$W^\mu$ agrees with the Noether current $J^\mu$  associated to the $U(1)$ global gauge symmetry  (up to an overall factor):  
\begin{align}
 J^\mu = iq \phi (D^\mu \phi)^* = - m^2  W^\mu  .
\end{align}
Since the Noether current $J^\mu$ is conserved  $\partial_\mu J^\mu = 0$, the $W^\mu$ satisfies the (divergenceless) relation:
\begin{align}
 \partial_\mu W^\mu = 0 ,
\end{align}
which is regarded as the subsidiary condition for the massive field $W^\mu$.

The conserved Noether charge $Q$ is a generator of the  global $U(1)$ transformation:
\begin{align}
 \delta \phi(x) = [ i\theta Q, \phi(x) ]
= i\theta \int d^dy [J^0(y), \phi(x) ]
= i \theta q\phi(x) .
\end{align}
where 
$
 J^0 = iq \phi \Pi_\phi
$.
This is consistent with no SSB:
\begin{align}
 \langle 0| \delta \phi(x) |0 \rangle 
= i \theta q  \langle 0| \phi(x) |0 \rangle = 0 .
\end{align}

\section{Gauge-invariant Higgs mechanism in SU(2) case:  partial SSB}

We consider $G=SU(2)$ Yang-Mills-Higgs theory  with the \textbf{adjoint scalar field} $\bm{\phi}(x)$ described by the gauge-invariant Lagrangian:
\begin{align}
 \mathscr{L}_{\rm YMH} =& - \frac{1}{4} \mathscr{F}^{\mu\nu }(x) \cdot \mathscr{F}_{\mu\nu}(x) 
+\frac{1}{2 } ( \mathscr{D}^{\mu}[\mathscr{A}] \bm{\phi}(x) ) \cdot (\mathscr{D}_{\mu}[\mathscr{A}] \bm{\phi}(x) )
- V(\bm{\phi}(x) \cdot \bm{\phi}(x)) .
\label{SU2-YMH-1}
\end{align}
For the Lie-algebra $su(2)$ valued quantities $\mathscr{A}=\mathscr{A}^AT_A$ and $\mathscr{B}=\mathscr{B}^AT_A$, we use the notation: 
\begin{align}
\mathscr{A} \cdot \mathscr{B}:=2{\rm tr}(\mathscr{A}\mathscr{B}) = \mathscr{A}^A \mathscr{B}^B 2{\rm tr}(T_A T_B) = \mathscr{A}^A \mathscr{B}^A \quad (A=1,2,3) .
\end{align}
We assume that the \textit{adjoint scalar field $\bm{\phi}(x)=\phi^A(x)T_A$ has the fixed radial length}:
\begin{align}
   \bm{\phi}(x) \cdot \bm{\phi}(x) \equiv \bm{\phi}^A(x)   \bm{\phi}^A(x) = v^2  
  .
\label{SU2-YMH-1-constraint}
\end{align}
The Yang-Mills field $\mathscr{A}_\mu(x)=\mathscr{A}_\mu^A(x)T_A$ and   the  scalar field $\bm{\phi}(x)$ obey the gauge transformation:
\begin{align}
    \mathscr{A}_\mu(x) &\to U(x) \mathscr{A}_\mu(x) U^{-1}(x) + ig^{-1} U(x) \partial_\mu U^{-1}(x) ,
\nonumber\\ 
\bm{\phi}(x)  &\to 
 U(x) \bm{\phi}(x) U^{-1}(x)  
 ,
\quad U(x) \in G=SU(N)  .
  \label{gauge-transf}
\end{align}
Notice that $\bm{\phi}(x) \cdot \bm{\phi}(x)$ is a gauge-invariant combination.
The covariant derivative 
$\mathscr{D}_{\mu}[\mathscr{A}]   := \partial_{\mu}  - ig[ \mathscr{A}_{\mu},  \cdot ]$
transforms 
$\mathscr{D}_{\mu}[\mathscr{A}]  \to U(x) \mathscr{D}_{\mu}[\mathscr{A}]  U^{-1}(x)$. 
Therefore,  $\mathscr{L}_{\rm YMH}$ of (\ref{SU2-YMH-1}) with the constraint (\ref{SU2-YMH-1-constraint}) is  invariant under the local gauge transformation (\ref{gauge-transf}).

First, we recall the \textit{conventional description
for the Higgs mechanism}.  
If  $\bm{\phi}(x)$ acquires a non-vanishing VEV $\langle \bm{\phi}(x) \rangle=\langle \bm{\phi}  \rangle=\langle \bm{\phi}^A  \rangle T_A$,  then the covariant derivative of the scalar field reduces to
\begin{align}
 \mathscr{D}_{\mu}[\mathscr{A}]\phi (x)  
  := \partial_{\mu} \bm{\phi}(x) - ig[ \mathscr{A}_{\mu}(x),  \bm{\phi}(x) ] 
 \to 
- ig [   \mathscr{A}_{\mu}(x),  \langle \bm{\phi} \rangle ]  + ...,
\end{align}
 and the Lagrangian density reads
\begin{align}
\mathscr{L}_{\rm YMH}^{} \to 
& - \frac{1}{2} {\rm tr}_{G} \{ \mathscr{F}^{\mu\nu }(x) \mathscr{F}_{\mu\nu}(x) \} 
- g^2 {\rm tr}_{G} \{  [\mathscr{A}^{\mu}(x), \langle \bm{\phi} \rangle ]  [\mathscr{A}_{\mu}(x), \langle \bm{\phi} \rangle ] \}   + ...   
\nonumber\\ 
=& - \frac{1}{2} {\rm tr}_{G} \{ \mathscr{F}^{\mu\nu }(x) \mathscr{F}_{\mu\nu}(x) \} 
- g^2 {\rm tr}_{G} \{  [T_A, \langle \bm{\phi} \rangle ]  [T_B, \langle \bm{\phi} \rangle ] \}  \mathscr{A}^{\mu A}(x)  \mathscr{A}_{\mu }^B(x)   . 
\label{VEV-L}
\end{align}
To \textit{break spontaneously  the local continuous gauge symmetry  $G=SU(2)$ by realizing the   non-vanishing VEV $\langle \bm{\phi} \rangle$ of the scalar field $\bm\phi$}, we choose  the \textbf{unitary gauge} in which  $\bm{\phi}(x)$ is pointed to a specific direction $\bm{\phi}(x) \to \bm{\phi}_{\infty}$ uniformly over the spacetime. 

This procedure does not completely  break the original gauge symmetry $G$. Indeed, there may exist a subgroup $H$ of $G$ such that  $\bm{\phi}_{\infty}$ does not change under the  local $H$ gauge transformation.
This is the \textit{partial SSB} $G \to H$:  
the mass is provided  for the coset $G/H$ (broken parts), while the mass  is not supplied for the $H$-commutative part of   $\mathscr{A}_{\mu}$: 
\begin{align}
\mathscr{L}_{\rm YMH} \to   - \frac{1}{2} {\rm tr}_{G} \{ \mathscr{F}^{\mu\nu }(x) \mathscr{F}_{\mu\nu}(x) \}
- (gv)^2 {\rm tr}_{\textbf{G/H}} \{  \mathscr{A}^{\mu}(x)  \mathscr{A}_{\mu}(x)  \} . 
\label{massG/H}
\end{align}
After the partial SSB, therefore, the resulting theory is a gauge theory with the \textbf{residual gauge group} $H$.
For $G=SU(2)$, by taking the usual \textbf{unitary gauge}: 
\begin{align}
\langle \bm{\phi}_{\infty} \rangle = v T_3, 
 \quad {\rm or} \quad 
\langle {\phi}^A_{\infty} \rangle  = v \delta^{A3} ,
   \label{unitary-gauge}
\end{align}
the kinetic term generates the mass term:
\begin{align}
& - g^2 v^2 {\rm tr}_{G} \{  [T_A, T_3 ]  [T_B, T_3 ] \} \mathscr{A}^{\mu A}  \mathscr{A}_{\mu}^B  
=  
\frac{1}{2} (g v)^2(\mathscr{A}^{\mu 1}  \mathscr{A}_{\mu}^{1}  + \mathscr{A}^{\mu 2}  \mathscr{A}_{\mu}^{2}  )
  . 
\end{align}
The off-diagonal gluons $\mathscr{A}_{\mu}^{a}$ $(a=1,2)$ 
acquire the same mass $M_{W} :=gv$, while the diagonal gluon $\mathscr{A}_{\mu}^{3}$ remains massless. 
Even after taking the unitary gauge (\ref{unitary-gauge}), $U(1)$ gauge symmetry described by $\mathscr{A}_{\mu}^{3}$ still remains as the residual local gauge symmetry $H=U(1)$, which leaves  $\bm{\phi}_{\infty}$   invariant (the local rotation around the axis  of the scalar field $\bm{\phi}_{\infty}$).
Thus, the SSB is sufficient for the Higgs mechanism to take place. 
But, it is not clear whether the SSB is necessary or not for the Higgs mechanism to work.
This description for the Higgs phenomenon depends on the specific gauge. 


Next, we give a  
\textbf{gauge-invariant (gauge-independent) description for the Higgs mechanism}, which does not rely on the SSB, see ref.\cite{Kondo16} for the details.
We construct a composite vector field $\mathscr{W}_\mu(x)$ from  $\mathscr{A}_\mu(x)$ and  $\bm{\phi}(x)$ by 
\begin{align}
 \mathscr{W}_\mu(x) 
:=   -ig^{-1}  
  [\hat{\bm{\phi}}(x), \mathscr{D}_\mu[\mathscr{A}]\hat{\bm{\phi}}(x) ]
 , 
\quad \hat{\bm{\phi}}(x):=\bm{\phi}(x)/v .
\label{W-def1}
\end{align}
We find  that the kinetic term of the Yang-Mills-Higgs model is identical to the ``mass term'' of the vector field $\mathscr{W}_\mu(x)$:
\begin{align}
 \frac{1}{2 }  \mathscr{D}^{\mu}[\mathscr{A}] \bm{\phi}(x) \cdot \mathscr{D}_{\mu}[\mathscr{A}] \bm{\phi}(x)  
=   \frac{1}{2 } M_{W}^2 \mathscr{W}^\mu(x) \cdot \mathscr{W}_\mu(x) , 
\quad  M_{W} :=  gv ,
\label{W-mass}
\end{align}
as far as  the constraint ($\bm{\phi} \cdot \bm{\phi}  =  1$)  is satisfied. 
This fact is shown explicitly for $G=SU(2)$. 
Remarkably, the above \textbf{``mass term''  of $\mathscr{W}_\mu$ is gauge invariant}, since $\mathscr{W}_\mu$  obeys  the adjoint gauge transformation: 
\begin{align}
  \mathscr{W}_\mu(x) \to U(x) \mathscr{W}_\mu(x) U^{-1}(x)
  .
\end{align}
Therefore,  \textbf{$\mathscr{W}_\mu$ becomes massive without breaking the original gauge symmetry.} 
The  SSB of gauge symmetry is not necessary for generating the mass of $\mathscr{W}_\mu$, since we do not need to choose a specific vacuum from all  possible degenerate ground states distinguished by the direction of $\bm{\phi}$.  
The above equation (\ref{W-def1}) for \textbf{$\mathscr{W}_\mu$ gives a gauge-independent definition of the massive gluon mode  in the operator level.}
The relation (\ref{W-def1}) is also \textit{independent from the parameterization of the scalar field}.

\begin{figure}[h]
\centering
\includegraphics[width=2.5cm]{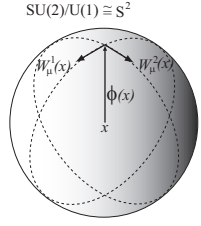}
\caption{The target space of the scalar field $\bm{\phi}(x)$ and the massive vector field $W_\mu^a(x)$ ($a=1,2$). The radially fixed scalar field $\bm{\phi}(x)$ takes the value on $S^2 \simeq SU(2)/U(1)$.  The vector field $W_\mu^a(x)$ must be orthogonal to $\phi(x)$. Therefore, each vector $W_\mu^a(x)$ lies in the  2-dimensional tangent space at $\bm{\phi}(x) \in S^2$.  
}
\label{fig:phi-W}   
\end{figure}

How is this description related to the conventional one?
Despite its appearance (\ref{W-def1}) of $\mathscr{W}_\mu$ obeying the adjoint gauge transformation, the independent internal degrees of freedom of the  field $\mathscr{W}_\mu=(\mathscr{W}_\mu^{A})$ $(A=1,2,3)$ is equal to  ${\rm dim}(G/H)=2$,  since 
\begin{align}
 \mathscr{W}_\mu(x) \cdot  \bm{\phi}(x) 
= 0 
 .
\label{W-def2}
\end{align}
Notice that this is a gauge-invariant statement. 
See Fig.~\ref{fig:phi-W}. 
Thus, \textbf{$\mathscr{W}_\mu(x)$ represent the massive modes corresponding to the coset space $G/H$ components} as expected. 

In fact, by taking the unitary gauge $\bm{\phi}(x) \to \bm{\phi}_{\infty}=v \hat{\bm{\phi}}_{\infty}$,   $\mathscr{W}_\mu$ reduces to 
\begin{align}
 \mathscr{W}_\mu(x) 
\to   -ig^{-1}  
  [\hat{\bm{\phi}}_{\infty}, \mathscr{D}_\mu[\mathscr{A}]\hat{\bm{\phi}}_{\infty} ] 
=&    
  [\hat{\bm{\phi}}_{\infty},   [  \hat{\bm{\phi}}_{\infty},  \mathscr{A}_\mu(x)  ]] 
=  \mathscr{A}_\mu(x) - (\mathscr{A}_\mu(x) \cdot \hat{\bm{\phi}}_{\infty})\hat{\bm{\phi}}_{\infty} 
 .
\end{align}
Then $\mathscr{W}_\mu$ agrees with the off-diagonal components for the specific choice $\hat{\bm{\phi}}_{\infty}^A=\delta^{A3}$:
\begin{align}
 \mathscr{W}_\mu^{A}(x) 
\to   \begin{cases}
 \mathscr{A}_\mu^{a}(x)  &(A=a=1,2) \\
 0 &(A=3)
\end{cases}  .
\end{align}
The constraint $\bm{\phi} \cdot \bm{\phi} =v^2$ represents the vacuum manifold in the target space of the scalar field $\bm{\phi}$.
The scalar field $\bm{\phi}$ subject to the constraint $\bm{\phi} \cdot \bm{\phi} =v^2$ is regarded as the \textbf{Nambu-Goldstone modes} living in the \textbf{flat direction} at the bottom of the potential $V(\phi)$, giving the  {degenerate lowest energy states}.
Therefore, the massive field $\mathscr{W}_\mu$ is   formed by combining the  {massless (would-be) Nambu-Goldstone modes} with the original massless Yang-Mills field $\mathscr{A}_\mu$. 
This corresponds to the conventional explanation that the gauge boson acquires the mass by absorbing the Nambu-Goldstone boson to appear  in association with the SSB.

For more details and implications for quark confinement \cite{KKSS15}, see the original paper \cite{Kondo16} and references cited there.

\section{Introducing the Higgs scalar modes}
\label{sec-1}

\noindent
We can introduce the Higgs scalar  mode $\rho(x)$ by removing the constraint
$
 \bm{\phi}(x) \cdot \bm{\phi}(x)  = v^2  
$.
We introduce a unit field $\hat{\bm\phi} (x)$ satisfying $\hat{\bm\phi}(x) \cdot \hat{\bm\phi}(x)=1$ to separate   $\rho$:  
\begin{align}
    \bm\phi(x) = h(x) \hat{\bm\phi} (x)
= [v+\rho(x)] \hat{\bm\phi} (x) .
\end{align}
Then the covariant derivative of $\bm{\phi}$ is given by
\begin{align}
   \mathscr{D}_{\mu}[\mathscr{A}] \bm\phi(x) 
= (\partial_\mu h (x)) \hat{\bm\phi}(x)   + h(x)  (\mathscr{D}_{\mu}[\mathscr{A}] \hat{\bm\phi}(x))   .
\end{align}
The kinetic term of $\bm{\phi}$ yields the mass term of $W_\mu$ and the kinetic term of $\rho$ with   interactions: 
\begin{align}
& \frac{1}{2} \mathscr{D}^{\mu}[\mathscr{A}] \bm{\phi}(x) \cdot \mathscr{D}_{\mu}[\mathscr{A}] \bm{\phi}(x)  
=  \frac{1}{2} \partial^\mu h(x) \partial_\mu h(x) + \frac{1}{2} h(x)^2  (\mathscr{D}^{\mu}[\mathscr{A}] \hat{\bm\phi}(x) \cdot \mathscr{D}_{\mu}[\mathscr{A}] \hat{\bm\phi}(x)) 
\nonumber\\
=& \frac{1}{2} \partial^\mu h(x) \partial_\mu h(x) +   \frac{1}{2}  \frac{h(x)^2}{v^2} (gv)^2 \mathscr{W}^\mu(x) \cdot \mathscr{W}_\mu(x) , 
\nonumber\\
=& \frac{1}{2} \partial^\mu \rho(x) \partial_\mu \rho(x) +   \frac{1}{2} M_{W}^2   \mathscr{W}^\mu(x) \cdot \mathscr{W}_\mu(x) +  g^2v   \rho(x) \mathscr{W}^\mu(x) \cdot \mathscr{W}_\mu(x) + ...
 .
\end{align}
The potential term $V$ of the scalar field $\bm{\phi}(x)$ is rewritten in terms of $\rho(x)$ from 
\begin{align}
 \bm{\phi}(x) \cdot \bm{\phi}(x) = h(x)^2 = [v+\rho(x)]^2 .
\end{align}

\section*{Acknowledgements}

This work is  supported by Grants-in-Aid for Scientific Research (C) No.15K05042 from the Japan Society for the Promotion of Science (JSPS).



\begin{thebibliography}{99}
\bibitem{Nambu}
Y.Nambu and G. Jona-Lasinio, 
Phys. Rev.\textbf{112}, 345
 (1961).


\bibitem{Goldstone}
J. Goldstone, 
Nuovo Cimento \textbf{19}, 154--164 (1961).
\\
J. Goldstone, A. Salam and S. Weinberg, 
Phys. Rev. \textbf{127}, 965
 (1962).


\bibitem{Higgs1}
P.W. Higgs,
Phys. Lett. \textbf{12},  132
 (1964).
\\
P.W. Higgs,
Phys. Rev. Lett.  \textbf{13},  508
 (1964).


\bibitem{Higgs2}
F. Englert and R. Brout,
Phys. Rev. Lett. \textbf{13}, 321
 (1964). 


\bibitem{Higgs3}
G.S. Guralnik, C.R. Hagen, and T.W.B. Kibble,
Phys. Rev. Lett. \textbf{13}, 585 (1964). 


\bibitem{YM54}
C.N. Yang and R.L. Mills,
Phys. Rev. {\bf 96}, 191
 (1954). 


\bibitem{Wilson74}
K. Wilson, 
Phys. Rev. D{\bf 10}, 2445
 (1974).
 

\bibitem{Elitzur75}
S. Elitzur, 
Phys. Rev. D\textbf{12}, 3978
 (1975).  
\\
G. F. De Angelis, D. de Falco, and F. Guerra, 
Phys.Rev. D\textbf{17}, 1624
 (1978).  


\bibitem{KK85}
T. Kennedy  and  C. King, 
Phys. Rev. Lett. \textbf{55}, 776
 (1985). 
Commun. Math. Phys. \textbf{104}, 327
 (1986).


\bibitem{BN86}
C. Borgs and F. Nill,
Commun. Math. Phys. \textbf{104}, 349
 (1986).
Phys. Lett. B\textbf{171}, 289
 (1986). 
Nucl. Phys. B\textbf{270}, 92
 (1986). 


\bibitem{FMS80}
J. Fr\"ohlich, G. Morchio, and F. Strocchi,
Phys. Lett. B\textbf{97}, 249
 (1980).  
Nucl. Phys. B\textbf{190}, 553
  (1981). 


\bibitem{GOZ04}
J. Greensite, S. Olejnik, and D. Zwanziger,
Phys. Rev. D\textbf{69}, 074506  (2004).
e-Print: hep-lat/0401003. 


\bibitem{CG08}
W. Caudy and J. Greensite,
Phys. Rev. D\textbf{78}, 025018  (2008).
e-Print: arXiv:0712.0999 [hep-lat]. 






\bibitem{Kugo81}
T. Kugo, 
Prog. Theor. Phys. \textbf{66}, 2249
 (1981). 


\bibitem{OS78}
K. Osterwalder and E. Seiler,
Ann. Phys. \textbf{110}, 440
 (1978).
\\
E. Seiler, 
Lect. Notes Phys. \textbf{159}, 1
 (1982).  


\bibitem{FS79}
E. Fradkin and S. Shenker,
Phys. Rev. D\textbf{19}, 3682
  (1979). 


\bibitem{tHooft80}
G. 't Hooft,
Which Topological Features of a Gauge Theory Can Be Responsible for Permanent Confinement? 
Lecture given at Cargese Summer Inst.1979, 
NATO Sci.Ser.B \textbf{59}, 17 (1980). 


\bibitem{Kondo16}
K.-I. Kondo,
Phys. Lett. B  {\bf 762}, 219--224 (2016).
e-Print: arXiv: 1606.06194 [hep-th]


\bibitem{KKSS15}
K.-I. Kondo, S. Kato, A. Shibata and T. Shinohara,
Phys. Rept. \textbf{579}, 1
 (2015).
arXiv:1409.1599 [hep-th].






\end{thebibliography}
\end{document}